# Designing plasmonic eigenstates for optical signal transmission in planar channel devices

Upkar Kumar,[1] Sviatlana Viarbitskaya,[2] Aurélien Cuche,[1] Christian Girard,[1] Sreenath Bolisetty,[3] Raffaele Mezzenga,[3] Gérard Colas des Francs,[2] Alexandre Bouhelier,[2] * Erik Dujardin[1] *

[1] CEMES CNRS UPR 8011, 29 rue J. Marvig, 31055 Toulouse, France.

[2] Laboratoire Interdisciplinaire Carnot de Bourgogne, CNRS UMR 6303, Université de Bourgogne Franche-Comté, 9 Av. A. Savary, 21000 Dijon, France.

[3] ETH Zurich, Department of Health Sciences and Technology, Schmelzberg-strasse 9, CH-8092 Zurich, Switzerland

**ABSTRACT.** On-chip optoelectronic and all-optical information processing paradigms require compact implementation of signal transfer for which nanoscale surface plasmons circuitry offers relevant solutions. This work demonstrates the directional signal transmittance mediated by 2D plasmonic eigenmodes supported by crystalline cavities. Channel devices comprising two mesoscopic triangular input and output ports and sustaining delocalized, higher-order plasmon resonances in the visible to infra-red range are shown to enable the controllable transmittance and routing between two confined entry and exit ports coupled over a distance exceeding 2 μm. The transmittance is attenuated by > 20dB upon rotating the incident linear polarization, thus offering a convenient switching mechanism. The optimal transmittance for a given operating wavelength depends on the geometrical design of the device that sets the spatial and spectral characteristic of the supporting delocalized mode. Our approach is highly versatile and opens the way to more complex information processing using pure plasmonic or hybrid nanophotonic architectures.

**Introduction**

Signal transfer at faster rates, longer distances, along narrower pathways, with reduced losses and a higher bandwidth are some crucial impediments to the improvement of any information processing technologies. Optical transfer in free space, dielectric waveguides or fibers offers the best prospects for several of these criteria except for sub-wavelength spatial confinement. Yet, this particular point is of paramount importance to interface optical signal with electronic information processing and has been partially addressed by the photon-plasmon conversion using noble metal circuitry.[1-6] Surface plasmon (SP) propagated in films,[7-8] waveguided in grooves[9-11] or stripes[12-14] has proven efficient in transmitting or even processing information to and from optical and optoelectronic planar devices.[15-17] The optoelectronic conversion has become more challenging as the size of electronic devices is reaching the nanoscale though recent advanced functionalities based upon particle chains[18-20] and atomic-size gaps[21-26] could allow the next level of downsizing. Notably, even at the nanoscale, the dominant approach to signal transmission consists in propagating along a selected wave vector or inside a one-dimensional (1D) waveguide, but lossy in- and out-coupling usually limit the overall transfer efficiency.[27-28] A further step is to design the spatial distribution of the in-plane plasmon modes to implement the in-, out-coupling as well as the directional transmission and routing functions.[20] The use of spatial and spectral properties of higher order plasmon modes in 2D crystalline Au cavities have recently been proposed as a generic approach to integrated optical logic gates and, further, information processing.[29] The implementation of plasmonic devices from the properties of 2D cavity modes pertain to the ability to tailor their size and shape.[30-32] While simple and usually symmetrical shapes can be obtained directly by chemical synthesis, the full potential of arbitrarily shaped crystalline cavities can, so far, only be attained by the physical or chemical etching of Au platelets which has already led, for example, to antennas with improved performances.[33-34] Using this approach, we have recently demonstrated that a small defect, such as a hole, could significantly alter the modal behavior of 2D cavities pointing out the potential for more complex modal design.[35]

Here, we report the directional signal transmittance and routing mediated by the rich and complex spatial and spectral variation of the plasmonic eigenmodes supported by 2D crystalline structures. The potential of multiple input-output devices with a deterministic transmission between one entry and one exit ports is demonstrated in a diabolo-like structures composed of a rectangular channel flanked by two mesoscale triangular pads that sustain higher order modes in the visible spectral range,[29, 36] as schematized in Figure 1. The size of the structure is intermediate between the supra-and sub-wavelength size ranges so as to combine long range delocalization and sharp field enhancement at the input and output ports that are therefore precisely defined. In Figure 1c, an in-plane near-field intensity map is calculated over the entire diabolo for an optical excitation consisting of a Gaussian laser beam impinging on the structure at normal incidence. The beam is kept fixed at one corner of the triangular input cavity. A clear remote response is observed at the output cavity as far as the diagonally opposed corner. The full map suggests that the remote response is mediated by the excitation of a mode bridging the input and output locations via the channel. Interestingly, the output signal is almost annihilated when the incident polarization is rotated by 90° (Fig. 1d). The same near-field transmittance simulations provide the spectral distribution of the delocalized plasmon modes (Fig. 1b). The simulated transmittance spectra are monitored in (O) when the diabolo is excited with horizontal or vertical polarizations in (I). Upon horizontal polarization excitation (black curve), the symmetrical diabolo exhibits a 100-nm wide transmittance peak culminating at 810 nm. When the polarization is flipped (red curve), the spectrum is completely attenuated in this spectral region. These simulations strongly suggest that the transmittance is obtained by populating polarization-dependent 2D delocalized modes. Indeed, by simply moving the excitation to the lower corner of the input cavity, the signal is redirected to the upper corner of the diabolo output cavity through a symmetrical and energetically degenerated set of delocalized eigenstates. We experimentally explore this concept and demonstrate the effective spectral and spatial engineering of higher order plasmon modes in channel devices comprising input and output cavities. These information transfer structures may comply with the upcoming demand of compact processing devices, such as logic gates,[28, 37] coupled to quantum emitters in integrated architectures aimed at transferring[6, 38-39] and manipulating[40-41] quantum information.

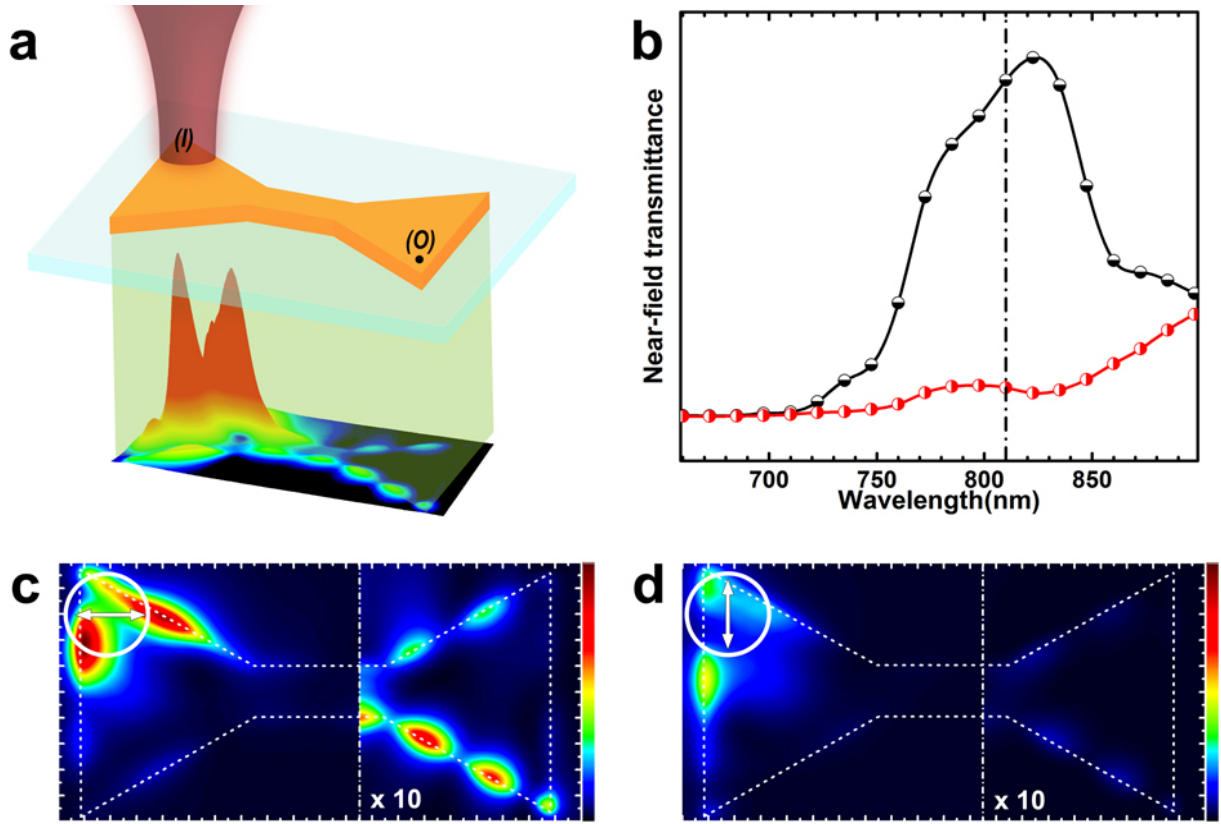

**Figure 1.** Modal transmittance in mesoscopic crystalline gold, diabolo-shaped devices. **(a)** Scheme for the numerical computation of the transmittance through the eigenstates of a channel device connecting triangular cavities. A focused pulsed laser beam is parked on the top left corner of the device (I). The transmitted near-field is calculated in all locations along the entire structure. Particular attention is paid to the farthest point (O) indicated by the black dot. The underlying map is a 3D rendering of panel (c). **(b)** Transmittance spectra calculated in (O) for a diabolo comprising 925-nm sided input and output triangular pads and a $500 \times 200$ nm channel excited with a linearly polarized Gaussian beam at 0° (black circles) and 90° (red circles). **(c, d)** Corresponding simulated near-field transmittance maps. The white circles indicate the position of the realistic 300-nm diameter, Gaussian excitation spot ($\lambda_{exc}$ = 810 nm, dash-dotted line in (b)), which is linearly polarized at (b) 0° and (c) 90° (See Suppl. Info for details). The color scale is multiplied by 10 to emphasize the transmission through the channel toward the exit port (right side of the device),

**Results and Discussion**

Focused ion beam (FIB) milling of crystalline gold microplatelets is used to design a structure supporting cavity modes that can be spatially and spectrally optimized to perform the desired transmission and routing functions from a chosen input port to a specific output location (Fig. 2a). This nanofabrication method provides plasmonic nanocircuits and devices with superior structural and optical quality.[33] The microplatelets are synthesized by a simple green reducing process reported in reference [42]. The colloidal suspension is drop-casted onto cross-marked glass coverslip and thoroughly washed. The microplatelets are natively stabilized by strongly adsorbed β-lactoglobulin amyloid fibrils, which are removed by three successive $O_2$-plasma treatments. We conducted scanning electron microscopy (SEM) and atomic force microscopy (AFM) to assess the scarce density of colloidal objects on the surface in order to address single objects. The in-plane dimensions (~ 2-3 μm) and thickness (20-40 nm) of the targeted native hexagonal microplatelets are recorded individually (Fig. 2b). Gallium ion FIB milling is performed, first, to etch out the diabolo shape (Fig. 2c) and then to remove the peripheral fragments of the mother microplatelet. The milling protocol is optimized using a pattern generator and nanolithography software to reduce edge amorphisation, metal re-deposition and gallium contamination. The typical side length of the equilateral triangular pads ranges from 600 to 1000 nm, which is comparable to chemically synthesized prismatic cavities, and the connecting channel is 500 nm long and 200 nm wide (See Experimental Section).[43]

Non-linear photoluminescence (nPL) microscopy is a convenient technique to map out the intensity of the local field inside plasmonic devices without resorting to slow and invasive near-field scanning probes.[29, 44] nPL imaging acts as a local probe that benefits from a higher spatial resolution and sensitivity conferred by the non-linear mechanism. In this work, the nPL of gold is excited with a 180-fs pulsed Ti:Sapphire laser focused in a diffraction-limited spot by a high numerical aperture objective (N.A. 1.49). The excitation spot diameter is about 300 nm and the mean power density at the sample is 20 mW.$\mu m^{-2}$. The power dependency of the non-linear emission signal collected in the 375–700 nm spectral range is quadratic and assimilated as such in corresponding simulations. Two different nonlinear optical mapping modes are implemented to obtain the modal landscape and the transmittance of the device. The first mode provides confocal maps reconstructed by raster scanning the device through the focal spot (Fig. S2).[43] The intensity of the detected nPL is proportional to the squared in-plane surface plasmon local density of states (SP-LDOS) and the maps reveal the spatial distribution of the eigenmodes of the 2D confined plasmons (Figs. S3 and S4).[29]

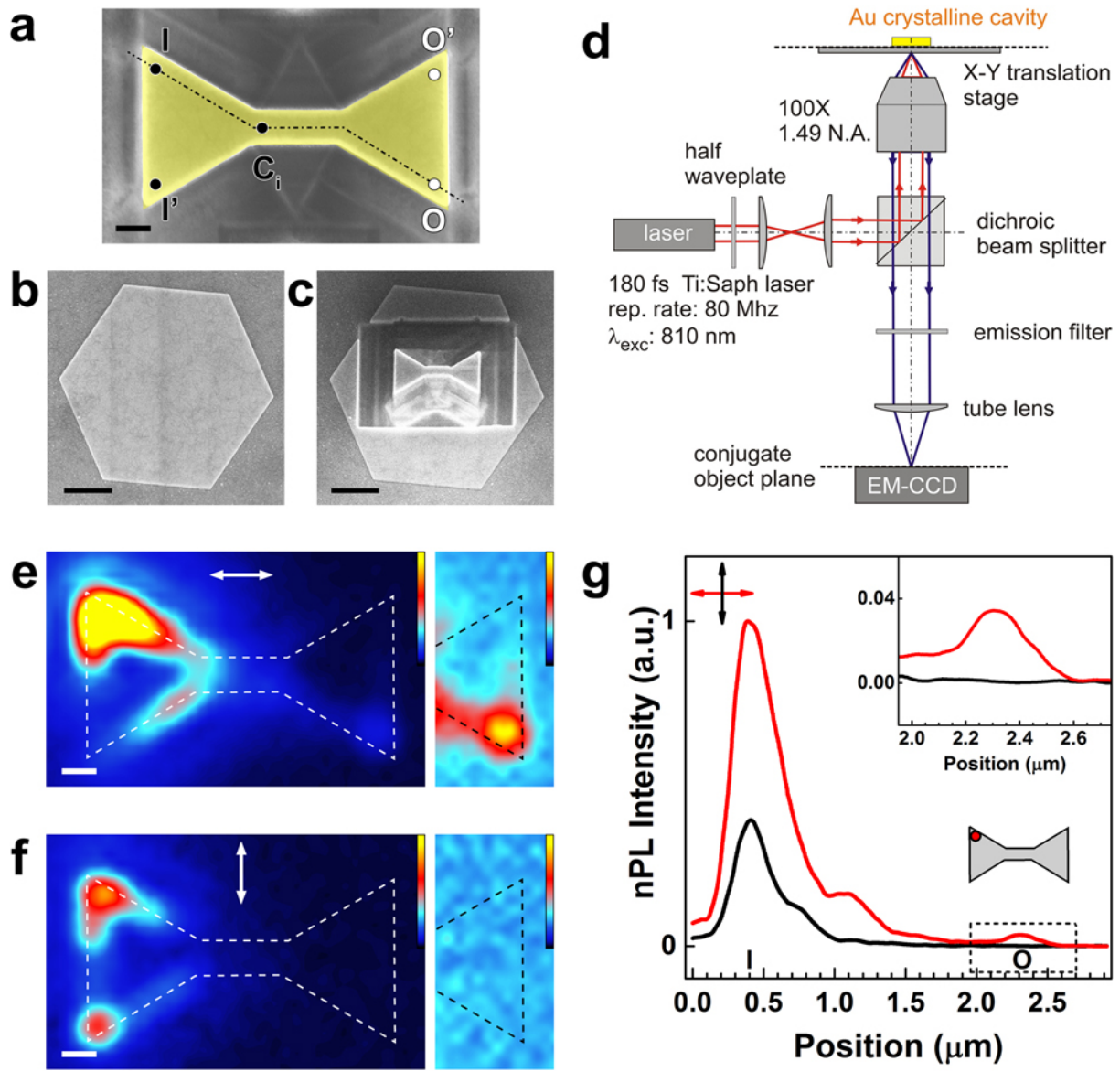

**Figure 2.** Experimental demonstration of the modal transmittance in diabolo-shaped devices **(a)** Scanning electron microscopy (SEM) image of a symmetrical diabolo-shaped structure. The sample is composed of two identical triangles (side length 925 nm) connected by a 500-nm long, 200-nm wide channel. (I), ($C_i$) and (O) indicated the specific input, channel input and output locations considered in this work. **(b)** SEM image of the pristine crystalline Au microplatelet deposited on ITO-coated glass coverslip prior to ion milling the sample shown in (a). **(c)** SEM image of the same microplatelet after nanopatterning the diabolo structure and prior to ion etching the unwanted peripheral parts of the starting platelet. **(d)** Schematics of the experimental set-up for wide-field non-linear photoluminescence (nPL) mapping. **(e, f)** Non-linear photoluminescence (nPL) wide-field CCD image of the diabolo structure shown in (a) when the 810-nm excitation is placed in (I) for an incident beam polarized at (d) 0°and (e) 90°. The insets on the right display the signal from the output right triangular area with a 10x

magnified color scale. **(g)** nPL profiles extracted from (d) and (e) along the path indicated by the dash-dotted line in (a). Inset: Magnified plot of the output (O) region corresponding to the dotted box in the main graph. Scale bars are (a, e, f) 200 nm and (b, c) 1 μm.

In the second imaging mode which is schematized in Figure 2d, the nPL response of the entire device is recorded for a fixed excitation position to visualize the plasmon-mediated signal transfer and distant nPL generation within the device.[45] Such wide-field images, displayed in Figures 2e and 2f, are obtained in leakage radiation (LR) imaging when the diabolo shown in Fig. 2a is excited in position (I) with a linearly polarized beam. In Figure 2e which is obtained for a 0° polarized excitation, nPL is not only observed as a local response coinciding with the excitation spot but also from several remote regions of the input triangle, the channel and even from the distant corner of the right triangle, where a localized bright spot can be observed in position (O). The nPL signal emitted in the substrate along the transfer path is collected in LR imaging mode with the high N. A. objective thus producing an image that can be qualitatively compared to the near-field response of the diabolo.[46-47] Indeed, the nPL distribution recorded in Figure 2e faithfully matches the transmittance map calculated in Figure 1c for the same excitation polarization direction, thus providing a compelling correlation between the transmission of the near-field from a fixed excitation and the remote production of nPL signal. The nPL images suggest that the pulsed excitation in (I) is indeed carried through the channel, as was recently observed in straight micrometer-long rods.[45] Yet, in contrast to long rods, which sustain a continuum of accessible resonances, the non-linear emission from the diabolo is not uniformly scattered from the remote (right) edge but it emerges from a very localized and specific location close to the corner, which coincides with an antinode of the SP-LDOS distribution (see Fig. 3a). This remote output spot could be used to further excite a collecting waveguide, forming an all-optical switch in an integrated nanophotonic platform.[37, 48] A comparison of the wide-field and confocal nPL maps confirms that the transmittance takes place between two locations of high SP-LDOS intensity, i.e. showing intense confocal nPL signal (Figs. S3 and S4). Indeed, the discrete modal structure of the diabolo is revealed in the computed maps of the projected in-plane SP-LDOS (Fig. 3 a-b) and the simulated nPL maps for an incident linearly polarized beam with a waist of 300 nm (Fig. 3 c-d). In particular, the four extremal hotspots in the corners of the triangular pads (Fig. 3c) are reminiscent of the ones observed in crystalline triangular prisms.[29] The input (I) and output (O) should be chosen in registry with these hotspots but the confocal nPL maps do not provide the transmittance connectivity between them. The simulated SP-LDOS map in Fig. 3a suggests a connecting modal path enabling a transfer of plasmon information through the channel, when the SP-LDOS is projected along the horizontal linear polarization. The transmission configurations with horizontally polarized excitation (Figs. 1c, 2e and S5) reveal that the four corners of the diabolo are connected in pairs by two energetically degenerated and diagonally delocalized modes that mediate the effective transmittance between (I) and (O) only or, similarly, between (I') and (O') only as schematized in Figure 3e (See also Fig. S5). In the confocal nPL maps, the relative intensity of the spots is adjusted by rotating the polarization direction away from horizontal.[43] Hence, an excitation with a vertical polarization does not result in a bright nPL emission from the distal tip of the triangular cavity as previously observed in the case of isolated prisms of similar size,[29, 31] but in a strongly reduced emission from the proximal corner and channel (Fig. 3d) associated with a minimal SP-LDOS distribution within the channel (Fig. 3b). In Figure 2f, this polarization configuration for the beam parked in input (I) markedly alters the intensity distribution transmitted in output (O) wherefrom no nPL emission is recorded. This is again in very good agreement with the simulated near-field transmittance map shown in Figure 1d. The effective signal transmittance for a horizontally polarized excitation and its variation upon rotating the polarization by 90° are further compared in Figure 2g by plotting the intensity profiles along the same I-$C_i$-O dotted path in Fig. 2a. While the confocal emission intensity in

(I) is only reduced by 55%, the transmittance modulation ON/OFF ratio in (O) reaches 130 i.e. a power attenuation of 21 dB.

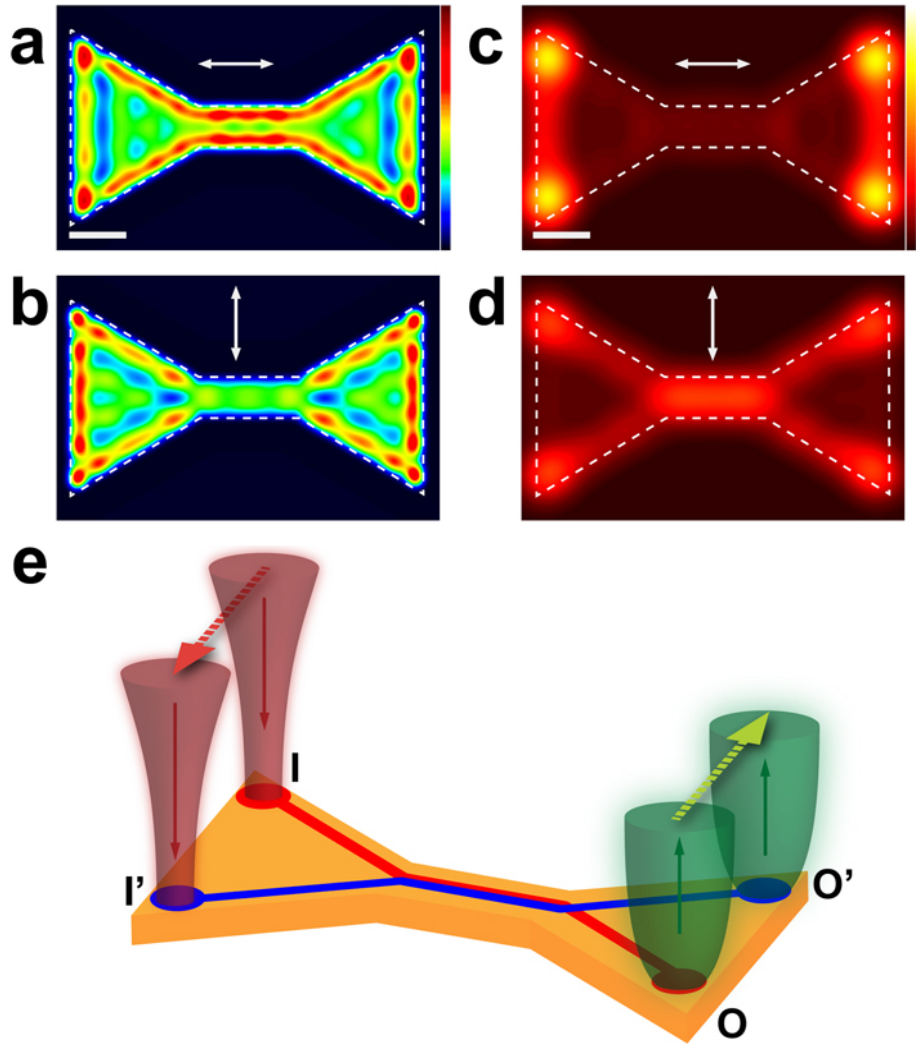

**Figure 3.** Plasmon modal properties of the mesoscale diabolo structure. **(a-b)** Partial in-plane SP-LDOS maps calculated for the symmetrical diabolo structure shown in Fig. 2a at λ = 810 nm and projected along linear polarization at (a) 0°, (b) 90°. **(c-d)** Simulated confocal nPL maps for a λ = 810 nm, linearly polarized excitation at (e) 0°, (f) 90°. The common scale and color bars are shown in (a) and (c) is 200 nm. (e) Schematic illustrating the link between the four symmetrical hotposts in the SPLDOS (a) and nPL (b) maps and the transmittance, through two diagonally delocalized degenerated plasmon modes (blue and red lines), from one

hotspot to the diagonally opposed corner where the nPL is remotely produced, as shown in Figs. 1 and 2.

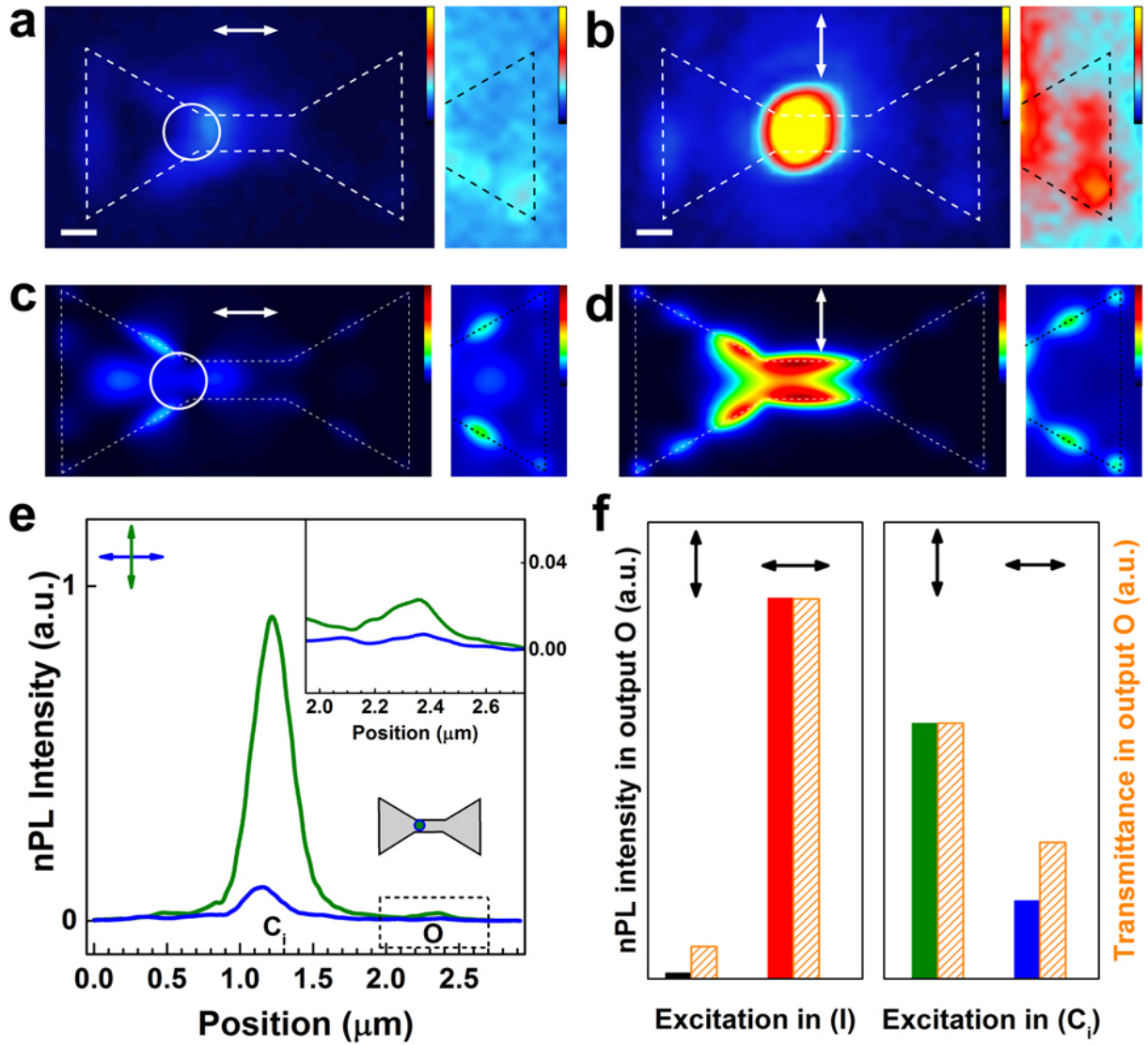

**Figure 4.** Direct channel excitation. **(a, b)** Non-linear photoluminescence wide-field CCD image of the diabolo structure shown in Fig. 1b when the excitation is placed at the channel input ($C_i$, white circle) for an incident beam polarized at (a) 0°and (b) 90°. The insets on the right display the signal from the right triangular area with a 10x magnified color scale. Scale bars are 200 nm. **(c,d)** Simulated near-field SP transmittance maps corresponding to (a) and (b) respectively. A realistic discretized 2D model of the sample and a Gaussian beam (300-nm

diameter, $\lambda_{exc}$ = 810 nm) are used. (**e**) Non-linear photoluminescence profiles extracted from (a) and (b) along the path indicated by the dash-dotted line in Fig. 2a. Inset: Magnified plot of the output (O) region corresponding to the dotted box in the main graph. (**f**) Histograms comparing the experimental non-linear luminescence intensity at the output O (full colored bars) with the near-field SP transmittance (striped orange bars) when the excitation beam is placed in either input I (left panel) or channel input Ci (right panel), for both 0° and 90° incident polarizations. The display of the SP transmittance is adjusted in each panel to match the largest experimental output of a given excitation location.

Figures 2 and 3 indicate that the transmittance is enabled by a spatial and spectral excitation of plasmonic eigen-modes of the diabolo. To further confirm the mode-mediated signal transfer, we considered exciting the structure in other locations while monitoring the non-linear emission at the output port (O). For example, in Figure 4, the excitation beam is parked at the left entrance of the channel (channel input point $C_i$, in Fig. 2a) which results in a striking change of behavior compared to the excitation in (I). A horizontally polarized excitation does not produce any delocalized nPL emission from the right pad (Fig. 4a) while a vertical polarization generates both an intense emission at the excitation point and a remote nPL spot in (O) as seen in the inset of Figure 4b. The corresponding simulated near-field SP transmittance maps are presented in Figures 4c and 4d for horizontally and vertically polarized excitation in $C_i$, with insets showing a 10x magnified intensity maps of the output regions. Both the relative intensity distributions and the spatial emission patterns observed in the nPL images are faithfully accounted for in the transmittance maps. The change of transmittance is clearly observed in the experimental intensity I-$C_i$-O profiles plotted in Figure 4e. A 90° polarization flip is accompanied by a 9.0-fold increase of the confocal emission but only induces an ON/OFF ratio of 3.2, or 5 dB, in spite of the ($C_i$)-(O) geometric distance (~ 1.1 μm) being about half the (I)-(O) distance (~ 1.9 μm). Significant dissipation effects that would limit the plasmon propagation can thus be excluded. In light of Figure 3, $C_i$ appears as a location with low density of states, for a 0° polarized excitation, which makes the energy in-coupling less effective and so results in a reduced transmittance efficiency in Fig. 4a. For a 90° polarized excitation in $C_i$, the in-coupling is slightly higher as indicated by a larger confocal response in $C_i$ seen both in Fig. 3d and in Fig. 4b, which coincides a significant nPL response in output port O in Fig. 4b. Interestingly, the polarization dependence of the transmittance when directly exciting the entrance of the channel at ($C_i$) is orthogonal to the case of a beam impinging in (I). This counter-intuitive behavior indicates that the channel cannot be considered as a rod nor the diabolo as the simple juxtaposition of two triangular and a rod-shaped sub-structures. The diabolo sustains specific resonances, the spatial distribution of which specifically enables the transport of plasmon information from a high SP-LDOS area to another via paths of finite SPLDOS present in the channel. This is precisely modelled in our plasmon transmittance simulation tool, in which the effective transmittance depends on the existence or not of a phase matching throughout the device (See Supporting Information, SI).[19] In Figure 4f, the transmittance of the device calculated for inputs (I) or ($C_i$) and output (O) is compared to the corresponding experimental nPL intensity measured in image plane maps for the two polarization configurations. The transmittance calculations faithfully follow the relative intensity of the output nPL intensity, which is maximized for the horizontal polarization when exciting in (I) and by the vertical polarization when exciting in ($C_i$). Moreover, for a given input location, the output transmittance in (O) in Figure 4f match the relative variation of the confocal nPL intensities in Figures 3 and S4 suggesting that the major contribution to the transmittance variation comes from the in-coupling efficiency. Indeed, a lower confocal nPL intensity in ($C_i$) is measured for the horizontal polarization (Figs. 3c and S4i) than for the vertical polarization (Figs. 3d and S4j), while the opposite is observed in (I). The efficiency to generate a non-linear output varies in other locations and is provided by the confocal nPL maps in Figures 3e-f and S4e-l. Some minor relative differences between the simulated and experimental transmittances are ascribed to the fact that our transmission model computes the linear near-field rather than the non-linear emission and does not account for the full complexity of the out-coupling to the far field and nPL collection, which may vary from one excited modes and for a particular polarization to another.

The operation of the transfer devices at a given energy relies on the spatial and spectral tailoring of a supporting delocalized mode that simultaneously enables an effective transmission and localized in- and out-coupling. This spatial and spectral matching of the cavities can be altered by varying the global geometry of the device. For instance, the symmetry breaking created in bowtie-like antennas comprising two mesoscale triangular prisms with different sizes was shown to have a marked effect on the local field distribution and spectral response.[36] We have fabricated non-symmetrical diabolos by changing the relative sizes of the triangular pads while keeping the same channel geometry. In Figure 5, we examine the case of a structure made of a 640-nm left side triangular pad connected to a 730-nm right side triangular pad (Fig. 5a). The total in-plane SPLDOS plotted in Figure 5b highlights the non-symmetrical modal structure with a predominant m=3 mode on the left side and a predominant m=5 mode on the right side. When the diabolo is excited at input (I) with a linearly polarized beam along the horizontal or vertical direction, no output is observed in (O) as shown in Figures 5c and 5d. The series of experimental nPL profiles in Figure 5g demonstrates that the transmittance along the I-$C_i$-O path remains inhibited for all polarization directions. The incident excitation propagates as far as the mid-channel area but is completely suppressed in the right side of the device. As the incident polarization direction is rotated, one can notice that the intensity of the nPL is uniformly affected over the entire structure, in striking contrast to the periodic intensity redistribution observed with the symmetrical diabolo for a similar sequence (See details in SI and Figure S6). The reshaping of the diabolo keeps the system off-resonance and does not allow for a signal transmission through the channel. We confirm the transmittance suppression by calculating the near-field SP transmittance maps for horizontally (Fig. 5e) and vertically (Fig. 5f) polarized excitation in (I). Once again, a very good match with the corresponding experimental maps is obtained. In the excitation regions, the confined hotspot and the extension along the edges observed in Figure 5c are matched by the two intense lobes in Figure 5e, while the weak and diffuse signal around (I) in Figure 5d is also obtained in the simulated Figure 5f. Furthermore, in the output regions, both polarization configurations yield a weak signal, which is marginally more intense in Figure 5e (horizontal polarization) than in Figure 5f (vertical polarization), in agreement with the weak transmittance measured in Figures 5g and S6. The spatially modulated, long-range transmittance monitored in the nPL images merely reveal in space the spectral match or mismatch of the device with the excitation conditions. In Figure 5h, the smaller non-symmetrical diabolo exhibits a sharper and blue-shifted transmittance peak at 775 nm (black curve) compared to the 820-nm resonance of the larger symmetrical device upon horizontally polarized excitation (Fig. 1b). Additionally, a low energy peak beyond 850 nm is observed in the non-symmetrical diabolo for the orthogonal excitation (red curve). Both excitation configurations though show a vanishing transmittance intensity at 810 nm, which confirms the off-resonance response of the non-symmetrical geometry at this wavelength. Similarly, the off-resonance excitation of the symmetrical diabolo at 750 nm results in no detectable nPL and in marginal transmittance in simulated maps in the output port, as shown in SI. The spectral differences between the two diabolos are both related to their shape differences and relative sizes, the non-symmetrical one comprising smaller triangular pads features resonances that are blue-shifted compared to the ones of the symmetrical device. Yet the smaller size itself does not warrant a higher transmittance which can be suppressed when operated out of resonance.

**Conclusion**

In this work, we illustrate how the modal engineering in two dimensions can be harnessed to design and optimize the optical signal transmission in compact and single crystalline devices that could be docked to other information processing modules. We have shown that the structural design of 2D mesoscale resonators from crystalline gold platelets allows to spatially and spectrally engineer SP modes exhibiting both delocalized extension and strong spatial modulation. Diabolo structures composed of a transmission channel flanked by two triangular resonators are investigated in detail. The excitation of the diabolo structures at specific input sites with suitable wavelength and polarization results in a remote but confined response, therefore implementing an efficient multi input/output signal transmission and routing through a two-dimensional device. The transmission mechanism is related to the existence of a complex plasmonic modal landscape. The transmission and routing functions are demonstrated experimentally using a fixed excitation and a detection of the spatial extension of the response through the channel. Our experimental results are confirmed numerically using a dedicated near-field SP transmittance code with a realistic polarized Gaussian excitation. The performances of the transmission modulation upon polarization flipping at 810 nm show a large ON/OFF ratio exceeding 20dB power attenuation for the spectrally-matched symmetrical diabolo. Our modal design approach to engineer and modulate SP transmission in compact wavelength-size devices contribute to the recent strategies to embed active information processing functions into pure or hybrid plasmonic structures such as the insertion of phase changing material into gold transmission lines[49] or the ultrafast electrical switching of SP in indium tin oxide waveguides.[50-52] The diabolo structures examined in this work are merely one first example of the potential of modal engineering in pure plasmonic systems toward information processing and transfer in complex devices, which could be used to create new computing architectures for classical and quantum optical technology.[38-39, 53]

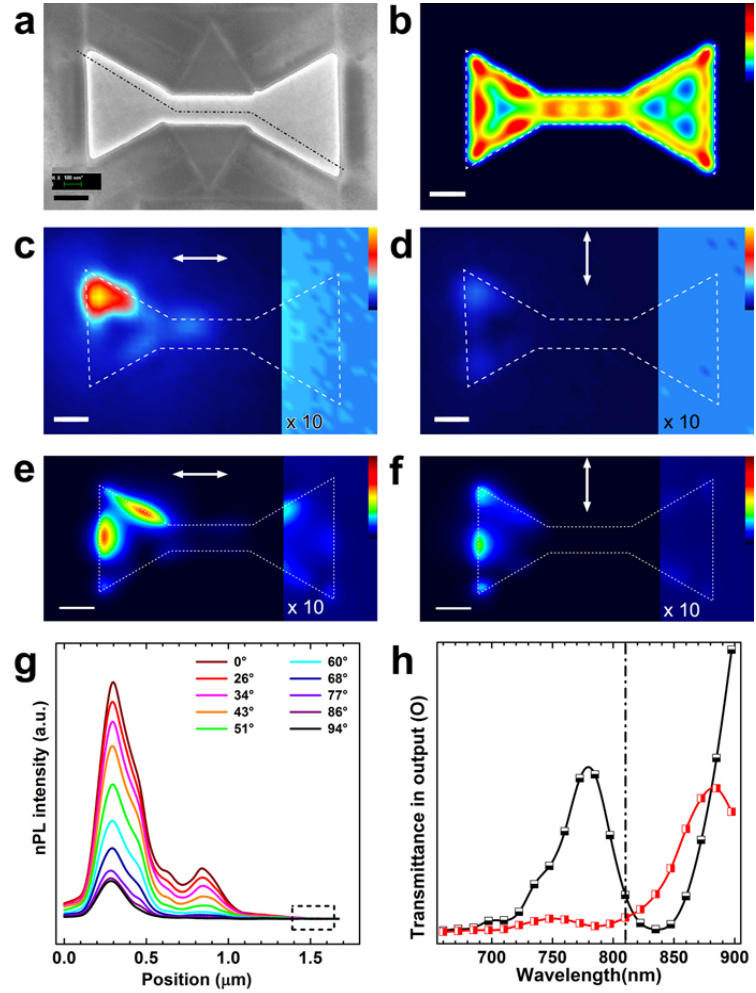

**Figure 5.** Transmission blockade by spectral mismatch in a non-symmetrical diabolo. **(a)** SEM image of a non-symmetrical diabolo structure composed of a small (630 nm side) and a large (730 nm side) triangles on the left (respectively right) of the 500 x 200 nm channel. **(b)** Full in-plane SP-LDOS map of the non-symmetrical diabolo shown in (a). **(c, d)** nPL images of the non-symmetrical diabolo structure shown in (a) when the excitation is placed in (I) for an incident beam polarized at (c) 0°and (d) 90°. **(e, f)** Simulated near-field SP transmittance maps corresponding to (c), (d) respectively. A realistic discretized 2D model of the sample and a Gaussian beam (300-nm diameter, $\lambda_{exc}$ = 810 nm) are used. **(g)** nPL profiles along the path indicated by the dash-dotted line in (a) for a series of polarization directions of the incident beam between 0° and 94°. The dashed box indicates the output O region. **(h)** Near-field

transmittance spectra calculated at readout (O) for an excitation in (I) with a linearly polarized Gaussian beam at 0° (black squares) and 90° (red squares). The dash-dotted line indicates the experimental excitation wavelength (810 nm). In (**c-f**) the color scale of the signal from the right triangular area is magnified 10 times. All scale bars are 200 nm.

### Experimental Section

**Sample preparation.** The crystalline gold microplatelets are synthesized using β-lactoglobulin protein fibrils according to the methods reported in references [42, 54]. The microplatelets studied here are prepared by initial mixing of 0.01M $HAuCl_4$ salt with 0.6 wt% of β-lactoglobulin fibrils and then heated at 55° C for 24 hours. The microplatelets are 20-40 nm in thickness and the typical in-plane dimensions are in the range of 1-15 μm. The colloidal suspension is drop-casted onto glass coverslips ($22 \times 22 \times 0.15$ mm) coated with a nominal 10-nm indium-tin oxide (ITO) layer. Cross-mark arrays are designed by photolithography and Au metallization. Organic residues are thoroughly washed with 10% aqueous ethanol solution and deionized water and the gold surface is decontaminated by three successive 5-min $O_2$ plasma treatments. The diabolo structures are fabricated by focused ion milling of hexagonal microplatelets with diameter of 3-5 μm using a gallium ion beam on a Zeiss 1540 XB dual beam microscope interfaced with a Raith Elphy Multibeam pattern generator. To ensure minimal substrate milling while thoroughly etching the gold, typical ion current of 1.0 pA and dose of 8000 μC.cm$^{-2}$ are used. The milling protocol consists in etching out the diabolo pattern first with minimal beam rastering along the diabolo edges. The peripheral portions of the starting hexagonal colloid are milled in a second step.

**Wide-field non-linear optical microscopy.** Wide-field non-linear optical microscopy is performed on an inverted microscope. A 180 fs pulsed Ti:Sapphire laser tuned at λ = 810 nm is focused in a diffraction-limited spot by a high numerical aperture objective (oil immersion, 100x, NA = 1.49). The full width half maximum (FWHM) spot diameter of the excitation beam is about 300 nm. The laser average power density at the sample is 20 mW.µm$^{-2}$. The linear polarization of the excitation beam is rotated by a half-wavelength plate inserted at the laser output. Non-linear photoluminescence (nPL) is collected through the same objective after filtering in the 375–700 nm spectral range from the backscattered fundamental beam by a dichroic beam splitter. Wide-field images of the signal emitted from the entire diabolo structure for a fixed excitation position are collected on a CCD camera placed in the conjugated object plane. Since the non-linear photoluminescence is quadratic with the excitation power, the experimental maps are normalized with the square of the excitation power measured at the back aperture of the objective.

**Simulations.** Numerical simulations of the local electromagnetic field, SP-LDOS and nPL signal were performed using home-made codes based on the 3D Green Dyadic Method as described in earlier reports.[44] The generalized propagator **K**(**r**,**r**˙,ω) is computed first that gives the total electromagnetic response of the complex metallic nanostructure under any arbitrary illumination field **E₀**(**R₀**, **r**, ω). The diabolo nanostructure is discretized into hexagonal lattice of cells. The local electromagnetic field inside the metallic nanostructure is then given by

$$E(\boldsymbol{R_0}, \boldsymbol{r}, \omega) = \int_V K(\boldsymbol{r}, \boldsymbol{r'}, \omega) \, . \, E_0(\boldsymbol{R_0}, \boldsymbol{r'}, \omega) d\boldsymbol{r'}$$

where **R₀** represents the center of the light beam. Simulations of confocal nPL intensity generated by a Gaussian beam excitation are obtained by integrating the square of local electric near-field intensity on the whole structure. The computed nPL maps are constructed by raster scanning the position of the sample through the focused Gaussian beam.[31, 44] The same formalism is used to calculate SPLDOS spectra and maps inside a metallic nanostructure.

The SP-LDOS transmittance calculations are performed with an excitation located at a fixed input co-ordinate (position vector r₁). The transmitted near-field intensity generated by the excited plasmonic field at a distal output co-ordinate (position vector r₂) is calculated. The energy transferred from input to output is obtained by solving the Dyson sequence equation which gives the Dyadic tensor S ($\mathbf{r_2}$, $\mathbf{r_1}$, $\omega$).

**Supporting Information**. Supporting information is available in the online version. Diabolo nanofabrication from single crystalline Au microplatelets (Fig. S1). Confocal nPL mapping (Fig. S2). SPLDOS, simulated and experimental confocal TPL maps (Figs. S3, S4). Signal routing through symmetrical degenerate plasmon modes (Fig. S5). Polarization dependency of the transmitted nPL intensity (Fig. S6). Calculation of near-field transmittance maps and spectra (Figs. S7, S8, S9). Resonant and non-resonant near-field transmittance maps and spectra in symmetrical diabolos (Fig. S10). References.

**Corresponding Authors**

* Email: Erik.Dujardin@cemes.fr

* Email : Alexandre.Bouhelier@u-bourgogne.fr

**Author Contributions**

ED, CG, AB designed the experiments. RM and SB synthesized the Au microplatelets. UK and ED conducted the sample preparation and nanofabrication steps. AB, SV implemented the experimental optical set-up. UK, SV, AC performed the optical experiments. CG, AC, UK, GCdF developed the simulations tools and performed all calculations. ED and UK processed the data. All authors contributed to the data analysis and manuscript writing. All authors have given approval to the final version of the manuscript.

**ACKNOWLEDGMENT**

This work was funded by the French Agence Nationale de la Recherche (Grant ANR-13-BS10-0007-ANR–PLACORE), the region of Burgundy under the PARI II Photcom and the European Research Council under the FP7/ 2007-2013 Grant Agreement No. 306772. The authors acknowledge the support of the massively parallel computing center CALMIP

(Toulouse, Fr).

**For table of contents only**

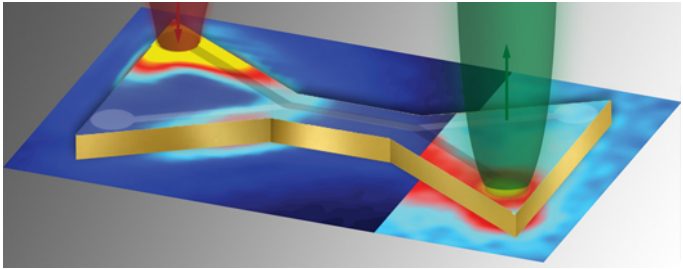

**Designing plasmonic eigenstates for optical signal transmission in planar channel devices.**

Upkar Kumar, Sviatlana Viarbitskaya, Aurélien Cuche, Christian Girard, Sreenath Bolisetty, Raffaele Mezzenga, Gérard Colas des Francs, Alexandre Bouhelier, Erik Dujardin.

Plasmonic modes in 2D gold cavities are designed to route non-linear signal from one input port to one output port and to modulate the transmitted power with the exciting polarization.

# Designing plasmonic eigenstates for optical signal transmission in planar channel devices

Upkar Kumar,[1] Sviatlana Viarbitskaya,[2] Aurélien Cuche,[1] Christian Girard,[1] Sreenath Bolisetty,[3] Raffaele Mezzenga,[3] Gérard Colas des Francs,[2] Alexandre Bouhelier,[2] * Erik Dujardin[1] *

[1] CEMES CNRS UPR 8011, 29 rue J. Marvig, 31055 Toulouse, France.

[2] Laboratoire Interdisciplinaire Carnot de Bourgogne, CNRS UMR 6303, Université de Bourgogne Franche-Comté, 9 Av. A. Savary, Dijon, France.

[3] ETH Zurich, Department of Health Sciences and Technology, Schmelzberg-strasse 9, CH-8092 Zurich, Switzerland

* Corresponding author: dujardin@cemes.fr

## Supporting information

## Table of contents

## S1. Diabolo nanofabrication from single crystalline Au microplatelets

The diabolo-shaped transmission devices were produced from crystalline gold microplates deposited on ITO-coated glass coverslip as described in the Methods section.

The large polydispersity of the colloidal suspension allowed to easily select hexagonal plates of adequate sizes (typically 3-5 micrometer diameter) that could be located optically or in the electron microscope with respect to labelled crossmarks (Fig. S1a and S1b).

The Ga ion milling is performed in two stages. First, the structure is defined and isolated by a sequence of irradiation boxes optimized to reduce the edge amorphisation and metal re-deposition (Fig. S1c). Then the peripheral areas of the starting platelet are removed leaving an isolated structure on the ITO/glass substrate (Fig. S1d).

Figures S1e to h show a series of several diabolo structures fabricated and studied in this work.

The fabrication protocol is also illustrated by the movie provided as a separate supplementary material.

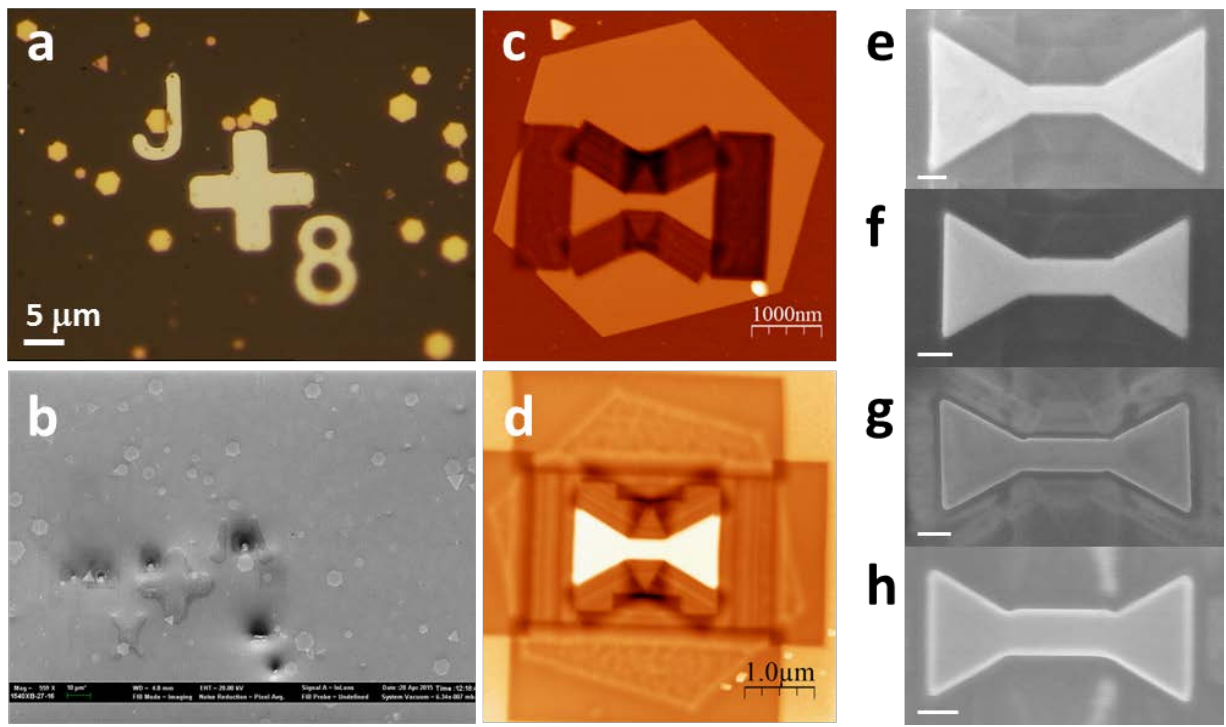

**Figure S1**: Diabolo nanofabrication. (a) Optical and (b) electronic images of the cross-marked ITO substrate bearing Au microplatelets. (c) AFM image recorded after the first FIB milling step and showing the diabolo structure defined inside a 5-$\mu$m diameter hexagonal platelet. (d) AFM image of the same sample after the second step dedicated to the removal of the peripheral Au platelet areas. (e-h) SEM images of several diabolo structures with different triangular pad sizes and channel dimensions. Scale bars are 200 nm.

## S2. Confocal nPL mapping

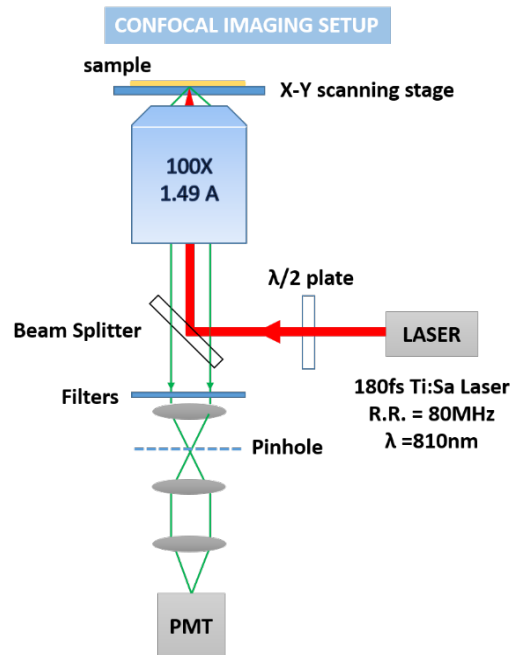

**Figure S2**: Schematics of the experimental set-up for confocal non-linear photoluminescence (nPL) mapping

Confocal non-linear optical microscopy is performed on an inverted microscope. A 180 fs pulsed Ti:Sapphire laser tuned at $\lambda$ = 810 nm is focused in a diffraction-limited spot by a high numerical aperture objective (oil immersion, 100x, NA = 1.49). The full width half maximum (FWHM) spot diameter of the excitation beam is about 300 nm. The laser average power density at the sample is 20 mW.$\mu$m$^{-2}$. The linear polarization of the excitation beam is rotated by a half-wavelength plate inserted at the laser output. Non-linear photoluminescence (nPL) is collected through the same objective, as the sample is raster scanned with a XY piezo stage (step size 25-50 nm), and filtered in the 375–700 nm spectral range from the backscattered fundamental beam by a dichroic beam splitter and a collection filter. Confocal maps are recorded on an avalanche photodiode. Since the non-linear photoluminescence is quadratic with the excitation power, the experimental maps are normalized with the square of the excitation power measured at the back aperture of the objective.

## S3 SP-LDOS, simulated and experimental confocal nPL maps

Confocal non-linear photoluminescence (nPL) mapping visualizes the spatial distribution of the surface plasmon local density of states (SP-LDOS) as demonstrated in refs [1-3]. Here we apply confocal nPL mapping to diabolo structures and we evidence modal behavior reminiscent to the one displayed by triangular nanoprisms. Experimental maps are compared to simulated nPL and SP-LDOS maps using the tools developed and described in our earlier work.

In particular, the SP-LDOS and nPL patterns generated by the symmetrical diabolo structure shown in Fig. S1e are displayed in Fig S3 for four different incident linear polarizations. Fig. S3a to S3d are the maps of the partial SP-LDOS at 810 nm projected on the polarization direction 0°, 90°, 120° and 150° with respect to horizontal. Our Green Dyadic Method code is used to simulate the realistic confocal nPL response (beam waist 300 nm) at 810 nm excitation for these four linear polarizations (Fig. S3e-S3h). These simulated maps can be compared to the experimental ones (Fig. S3i-S3l) acquired by raster scanning the 300-nm focused beam waist of the pulsed Ti : Sapphire laser operated at 810 nm and by collecting the non-linear luminescence signal (cut-off filter at 500 nm). Both experimental and simulated nPL maps patterns match and relative intensities on the apexes and in the channel recorded experimentally are well accounted for in the simulations.

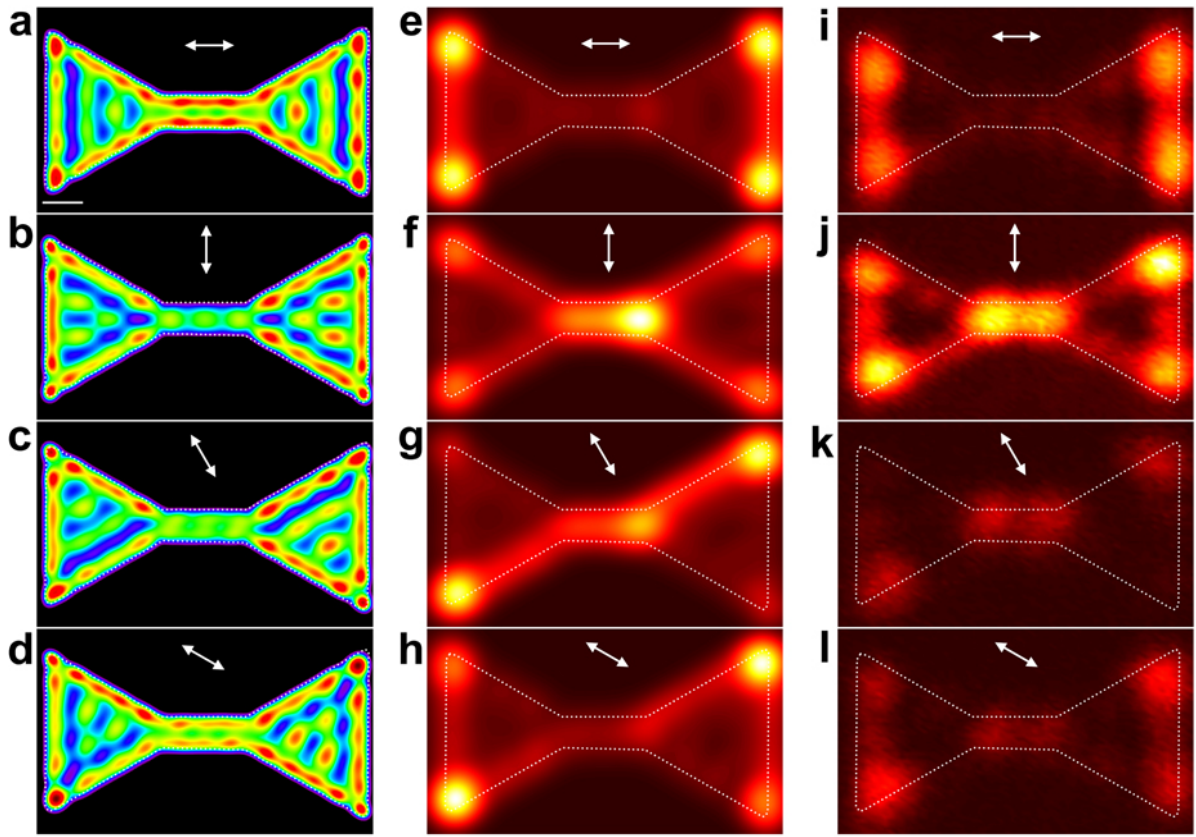

**Figure S3.** (a-d) Partial SP-LDOS maps at 810 nm, obtained for projection along the (a) 0°, (b) 90°, (c) 120°, (d) 150° polarization directions. (e-h) Simulated confocal nPL maps for an excitation at 810 nm, with a realistic beam waist of the incident Gaussian beam of 300 nm diameter. (i-l) Experimental confocal maps recorded for the same corresponding polarization directions. The sample is described in Fig. S1e.

The observation of intense and sharply localized luminescence spots at the outer apexes of the triangular pads, the intensity of which is modulated upon rotating the excitation polarization direction, is reminiscent of our earlier reports on the non-linear patterns produced by triangular nanoprisms. In particular, the longitudinal excitation (0° polarization; Figs. S3a, S3e and S3i) is associated with a minimal confocal signal in the channel and a maximal intensity emitted from the outer apexes. On the contrary, the transverse excitation (90° polarization; Figs. S3b, S3f and S3j) exhibits a reinforcement of the confocal emission localized in the channel. Intermediate projections, such as the one shown in Figs. S3c, S3g and S3k (polarization angle 120°) indicate that high nPL (i.e. high SP-LDOS) is distributed along a path connecting one apex on the left triangular pad to the opposite apex on the right pad, through the channel.

The exact same observations are made for the diabolo presented in Figure 2 of the main text. The SP-LDOS (Fig.S4a-d) and simulated nPL (Fig.S4e-h) match the corresponding experimental confocal nPL displayed in Fig. S4i-l. However, one can notice one bright spot on the lower edge of the right triangular pad is always visible in the confocal maps, irrespective of the excitation polarization direction, and is attributed to a small defect in the substrate near the diabolo. Such a defect does not affect transmittance when the diabolo is excited far from the defect (in position (I) for example) and the signal collected in leakage image plane mapping. This indicates that the defect might not be directly connected to the diabolo structure but within a spot radius when the confocal detection is recorded close to the channel entrance. Incidentally, this defect close but not on the diabolo may explain the small but real asymmetry observed in image plane maps recorded when exciting in Ci (Figs. 4a and 4b) that are not accounted for by the simulation on the realistic model (Figs. 4c and 4d). Indeed, when exciting in Ci, the defect may be excited alongside and generate a non-symmetrical input configuration.

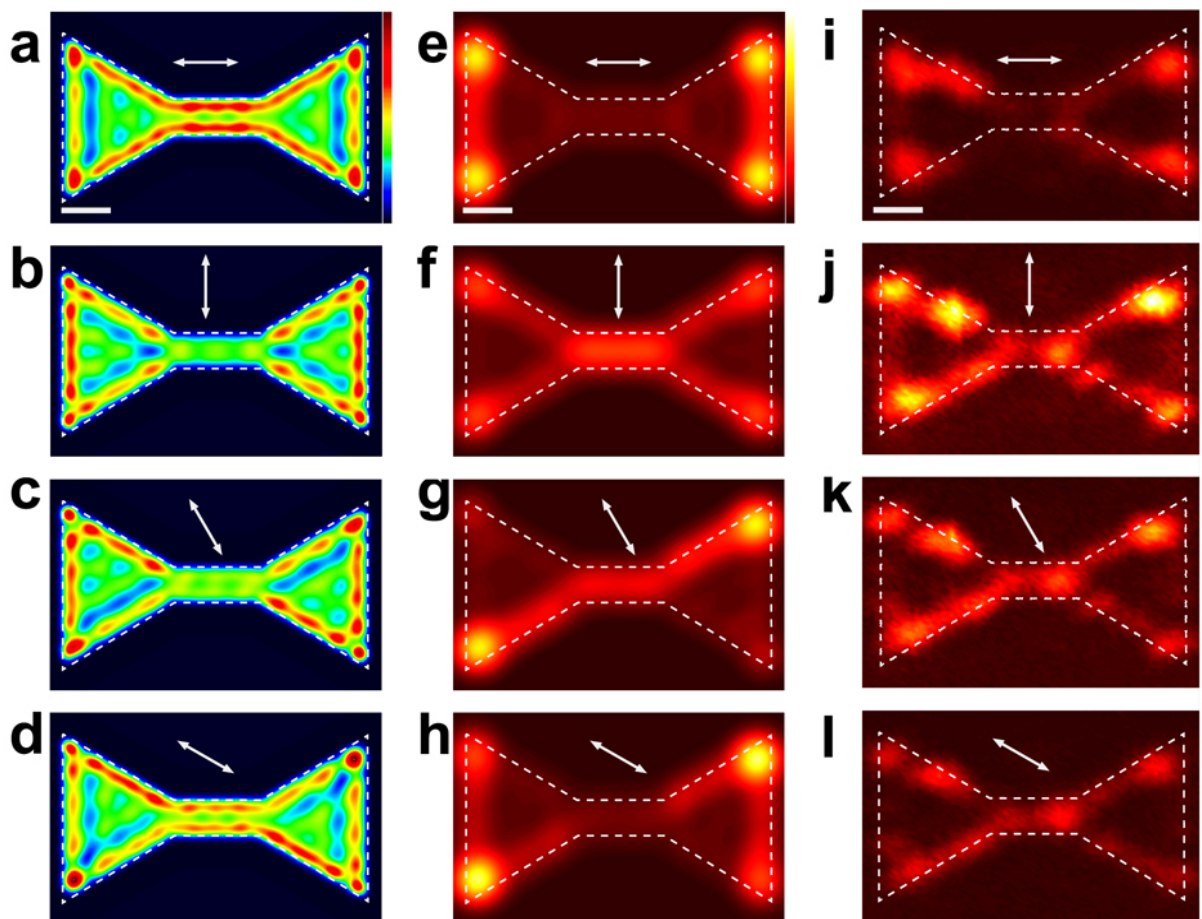

**Figure S4**. (a-d) Partial SP-LDOS maps at 810 nm, obtained for projection along the (a) 0°, (b) 90°, (c) 120°, (d) 150° polarization directions. (e-h) Simulated confocal nPL maps for an excitation at 810 nm, with a realistic beam waist of the incident Gaussian beam of 300 nm diameter. (i-l) Experimental confocal maps recorded for the same corresponding polarization directions. The sample is described in Fig. 2. Common scale bar is 200 nm.

## S4 Signal routing through symmetrical degenerate plasmon modes

The four SP-LDOS hot spots can be exploited to input and output optical signal as shown both numerically and experimentally in Figure S5 in complement to Figures 1 and 2. When the diabolo device is excited at position (I) (Fig. S5a) and (I') (Fig. S5a), the simulated transmission of the plasmonic signal to the right cavity of the diabolo device is seen at the diagonally opposed (O) and (O') respectively, for an excitation polarization along the channel. These identical but symmetrical transmission patterns establish the mirror symmetry of the optical signal transmission between paired SP-LDOS hotspots.

The experimental excitation in (O') (Fig. S5c) and (O) (Fig. S5d) results in localized outputs in (I') and (I) respectively and shows the same symmetrical coupling of the SP-LDOS hotspots (I)-(O) and (I')-(O') through degenerated modes linking these diagonally opposed apexes.

Incidentally, Fig. S5d demonstrates the reversible excitation at the position (O) with a horizontally polarized beam and output in (I) compared to Fig. 2d, where the excitation in (I) and the output in (O). The experimental wide field images and computed transmission images presented in Figure S10, thus confirm the reversibility and full symmetry of the optical transmission functionality of the diabolo device. The transmittance is mediated by a set of energetically degenerated plasmonic eigenstates that are all accounted for together in the corresponding confocal SP-LDOS maps (Fig. S4).

These results demonstrate that the diabolo structure acts as a routing transmission devices in which the choice of the excitation location, (I) or (I'), results in the non-linear emission from either (O) or (O') and vice-versa.

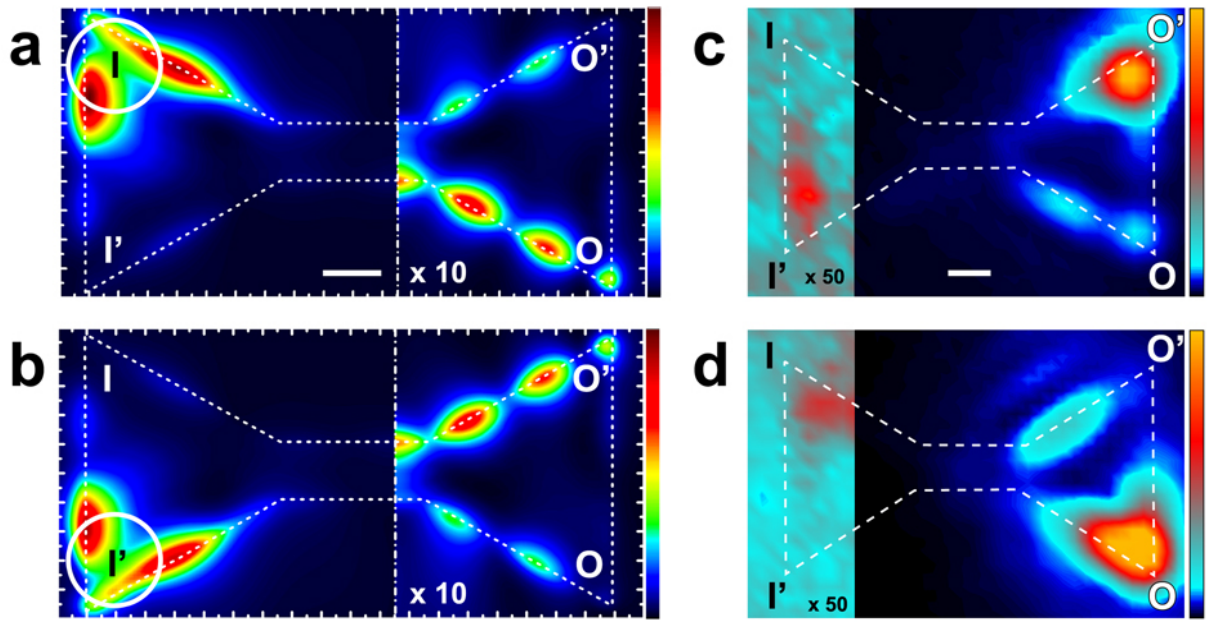

**Figure S5:** Symmetry and reversibility in the diabolo device: (a), (b) Simulated transmission maps for excitation at inputs (I) and (I'), respectively, with laser beam with horizontal polarization (0°). The wavelength used for the excitation is 810nm. The diabolo is composed of cavities of size 925nm connected by 200nm wide and 500nm long channel. The right side of the image has been enhanced in color scale by 10x for clear visibility of the transmitted signal. (c), (d) Experimentally recorded wide field nPL images for excitation at positions (O') and (O), respectively, with laser beam of wavelength 810nm. The diabolo used here is described in Fig. 2a. The left side of the cavity is color enhanced by a factor 10x.

**S5 Polarization dependency of the nPL intensity in image plane maps**

The transmittance through the diabolo structure is highly modulated by the excitation polarization direction (See Figs. 1, 2, 4). In Figure S6, we show the full polarization dependency of the signal recorded on image plane nPL maps at the location of the excitation beam (Input I), at the channel entrance (Ci) and at the readout output (O) for the symmetrical (Fig. S5a) and non-symmetrical (Fig. S5b) diabolos studied in the main text.

In Figure S6a, one observes that the nPL signal is alternatively maximum above each successive apex (the first being (I), the second one being the channel input Ci and the third one being the lower apex). Strikingly, the output (O) polarization dependency strictly matches the one in (Ci). The maximum intensity on (I), (Ci) and the lower apex are shifted by about 80°-90°. This graph suggest that the transmittance remains low for most polarization configurations until the channel is adequately excited – for angles close to 0° - and then the funneling as far as the the output (O) is set without further polarization shift (red and blue curves overlap). This behavior indicates the excitation of a resonance mode delocalized over the 2D diabolo structure that is characterized by a sharp polarization-dependent condition.

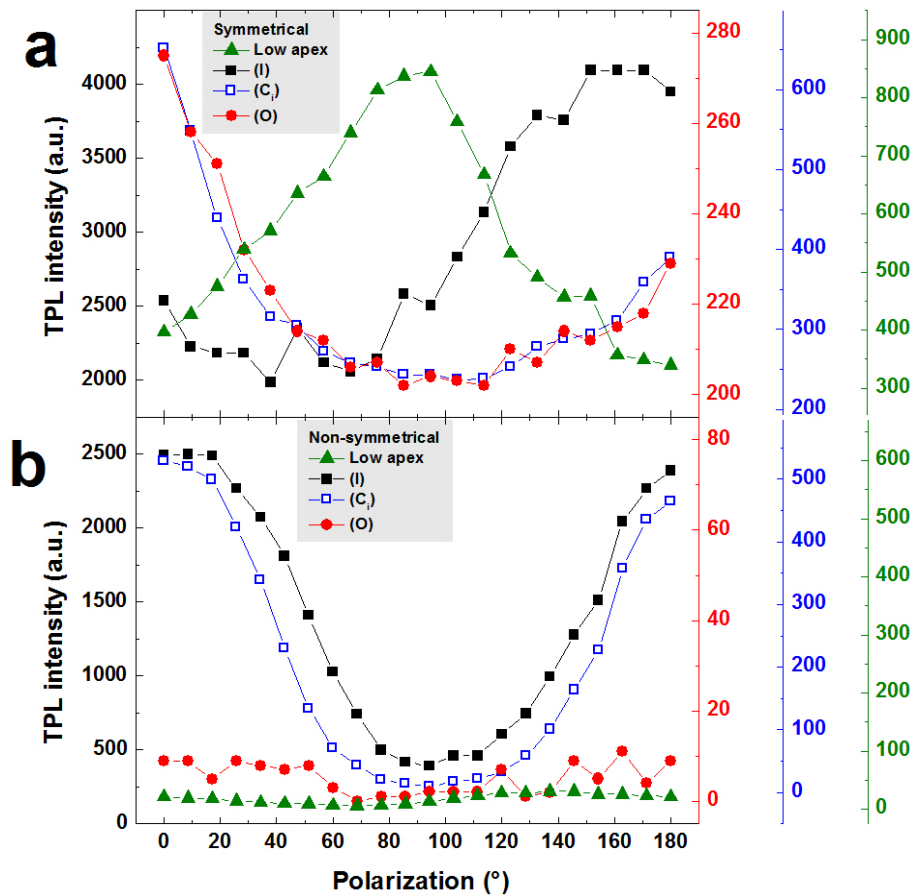

**Figure S6**. Evolution of the nPL intensity recorded in points I (black squares), $C_i$ (blue open squares), O (red dots) and the lower apex of the excited triangular pad (green triangles) from the image plane maps obtained for an excitation in (I) with the polarization direction between 0° and 180° with respect to the diabolo main axis. Panel (a) refers to the symmetrical diabolo shown in Fig. 2a and panel (b) refers to the non-symmetrical diabolo shown in Fig. 4a. Each dataset has its own color-coded Y-scale. Note that the amplitude of the scales for a given point are close to identical between panels (a) and (b) for relative comparison but the origins have been slightly shifted for clarity.

On the contrary, the transmittance intensity along the non-symmetrical diabolo shows a perfectly overlapping polarization evolution in (I) and (Ci) suggesting a global non-resonant modulation. The third apex and the output remain close to their minimal value with a very small modulation also in phase. When excited at 810 nm, the asymmetrical produces a transmitted luminescence that is spatially attenuated (See Fig 5) and modulated uniformly by the incident polarization.

The polarization dependency reported here reinforce the attribution of the effective transmittance to the excitation of a delocalized SP mode that display SP-LDOS intensity extrema at opposite apexes used as input (I) and output (O).

## S6 Calculation of near-field transmittance maps and spectra

Here, we describe the numerical tool based on the Green dyadic formalism that provides the computation of the electric field transfer between two remote locations in an arbitrary shaped metallic nanostructure as illustrated in Figure S7.

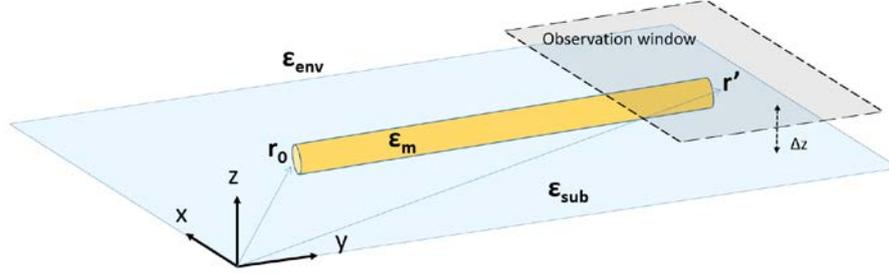

**Figure S7**. Schematic geometry of a plasmonic 1D channel, supported by a transparent dielectric substrate, and the associated observation window above the structure used for simulations. The size of this window can be adjusted to the dimensions of the structure.

In a metallic system where surface plasmons can be excited by a focused illumination $\mathbf{E_0}$, of angular frequency $\omega$ and positioned at $\mathbf{r_0}$, the expression of the electric field $\mathbf{E_{out}}$ at a given output location $\mathbf{r'}$ can be expressed as follows:[4,5]

$$E_{out}(\mathbf{r'}, \mathbf{r_0}, \omega) = \int_V K(\mathbf{r'}, \mathbf{r}, \omega).E_0(\mathbf{r}, \mathbf{r_0}, \omega)\, d\mathbf{r} \qquad (1)$$

where $V$ is the volume of the metallic object, and $K(\mathbf{r'}, \mathbf{r}, \omega)$ is the generalized field propagator that contains the entire response of the sample under any incident illumination:[4]

$$K(\mathbf{r'}, \mathbf{r}, \omega) = \delta(\mathbf{r} - \mathbf{r'}) + \chi(\mathbf{r}, \omega).S(\mathbf{r'}, \mathbf{r}, \omega) \qquad (2)$$

$S(\mathbf{r'}, \mathbf{r}, \omega)$ is the Green Dyadic tensor describing the whole system that includes the plasmonic structure and the dielectric substrate. In addition, the frequency-dependent optical response of the metallic structure is given by:

$$\chi(\mathbf{r}, \omega) = \frac{\varepsilon_m(r, \omega) - \varepsilon_{env}}{4\pi} \text{ in the metal,} \qquad (3)$$

and $\chi(\mathbf{r}, \omega) = 0$ outside of the metallic object.

In this work, we focused on the plasmon-mediated field transfer in the metallic structures. We therefore excluded the direct contribution of the incident illumination at the output location.[6] This can be done by removing the delta Dirac distribution in equation **(2)**:

$$K(\mathbf{r'}, \mathbf{r}, \omega) = \chi(\mathbf{r}, \omega).S(\mathbf{r'}, \mathbf{r}, \omega) \qquad (4)$$

The whole set of experimental data presented in this study has been acquired using a focused Gaussian illumination. Consequently, the illumination field $\mathbf{E_0}$ has been modelled as a Gaussian spot through an expansion in a plane waves in the wave vector domain:[7-10]

$$E_0(\mathbf{r}, \mathbf{r_0}, \omega) = \int_{-\sqrt{\varepsilon_{sub}}k_0}^{\sqrt{\varepsilon_{sub}}k_0} d\alpha \int_{-\sqrt{\varepsilon_{sub}k_0^2 - \alpha^2}}^{\sqrt{\varepsilon_{sub}k_0^2 - \alpha^2}} d\beta\, \zeta\, exp\left[-\frac{w_0^2(\alpha^2 + \beta^2)}{4}\right] exp[i\alpha(x - x_0) + i\beta(x - y_0) + ik_z(x - z_0)],$$

$$(5)$$

$k_0$ is the vacuum wave vector of the incident light and $\boldsymbol{r_0} = (x_0, y_0, z_0)$ defines the center of the excitation Gaussian spot in the Cartesian system of coordinates shown in Fig. S7. $\varepsilon_{sub}$ is the dielectric function of the substrate and the beam waist $w_0$ describes the lateral extension of this incident Gaussian beam. The integration in equation **(5)** is performed in the 2D reciprocal space defined by the vector $\boldsymbol{k_\parallel} = (\alpha, \beta)$. In order to take into account the realistic geometry of the system with a transparent substrate, the tangential components of the field vector $\zeta$ are expressed as follows:

$$\begin{pmatrix} \zeta_x \\ \zeta_y \end{pmatrix} = \mathrm{T} \begin{pmatrix} E_{0,x} \\ E_{0,y} \end{pmatrix}, \tag{6}$$

The transmission matrix T reads:

$$\mathrm{T} = \begin{pmatrix} (\tau_\parallel - \tau_\perp)\cos^2\delta + \tau_\perp & (\tau_\parallel - \tau_\perp)\cos\delta\sin\delta \\ (\tau_\parallel - \tau_\perp)\cos\delta\sin\delta & (\tau_\parallel - \tau_\perp)\sin^2\delta + \tau_\perp \end{pmatrix}, \tag{7}$$

$\tau_\parallel$ and $\tau_\perp$ are the Fresnel coefficients for the interface. Here, $\delta$ is an angle in the xy plane between the x-axis in Cartesian coordinates and the orientation of the planar component of the wave vector $\boldsymbol{k_\parallel}$. In equation **(5)**, the normal component $\zeta_z$ of $\boldsymbol{\zeta}$ is obtained using the following expression:

$$\zeta_z = -\left(\alpha\zeta_{inc,x} + \beta\zeta_{inc,y}\right)\Big/\left(\varepsilon_{sub}k_0^2 - \alpha^2 - \beta^2\right)^{1/2} \tag{8}$$

Finally, the electric intensity distributions generated in the simulated maps of the main manuscript and the figures S8 and S9 below take the following form:

$$I(\boldsymbol{r'}, \boldsymbol{r_0}, \omega) = |\boldsymbol{E}(\boldsymbol{r'}, \boldsymbol{r_0}, \omega)|^2 \tag{9}$$

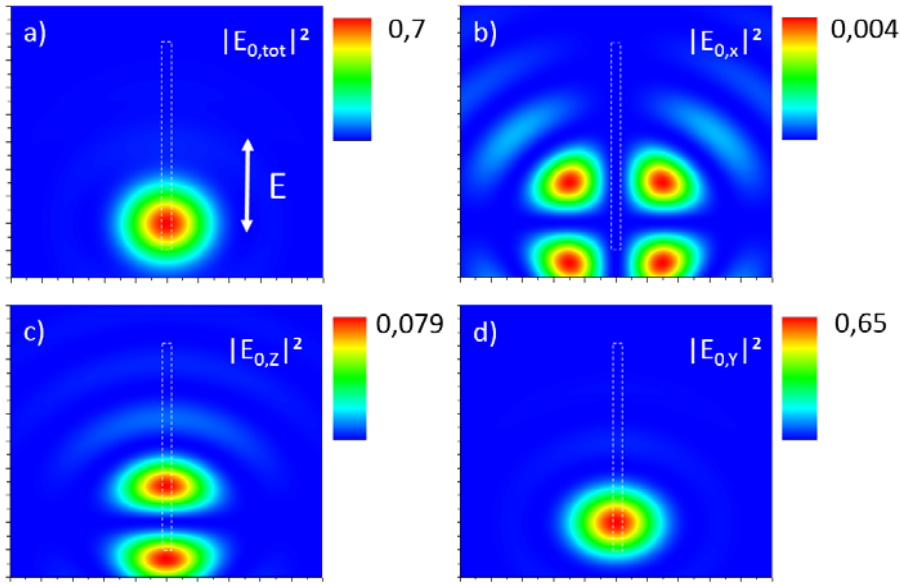

**Figure S8**. 2 x 2 µm² simulated images showing the normalized intensity distribution of the incident Gaussian illumination spot (a) Total normalized intensity for a polarization along the main axis of the wire. (b-d) Normalized intensity maps of the (b) x, (c) y and (d) z components of the field. The spot is placed at the location of one extremity of a 1.5-µm long, 50-nm diameter gold wire that is considered in Figure S8 and indicated here by the white dashed contour. In these simulations, the nanowire is not included.

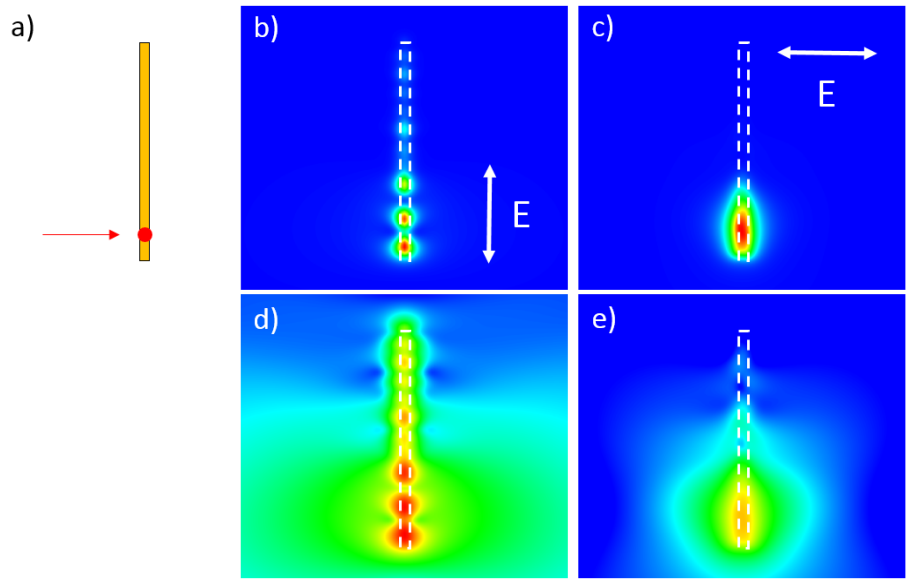

**Figure S9**. (a) Schematic view of the Gaussian excitation spot position (red dot) with respect to the 1.5 µm long gold wire. (b-c) Corresponding 2 x 2 µm² simulated images showing the intensity distribution of the total field propagating in the wire once illuminated by the Gaussian spot. (d-e) Similar to (b-c) with a logarithmic color scale. The incident polarization is indicated by the white arrows. The four maps are computed at a distance of 30 nm above the metallic wire.

## S7 Resonant and non-resonant near-field transmittance maps and spectra in symmetrical diabolos

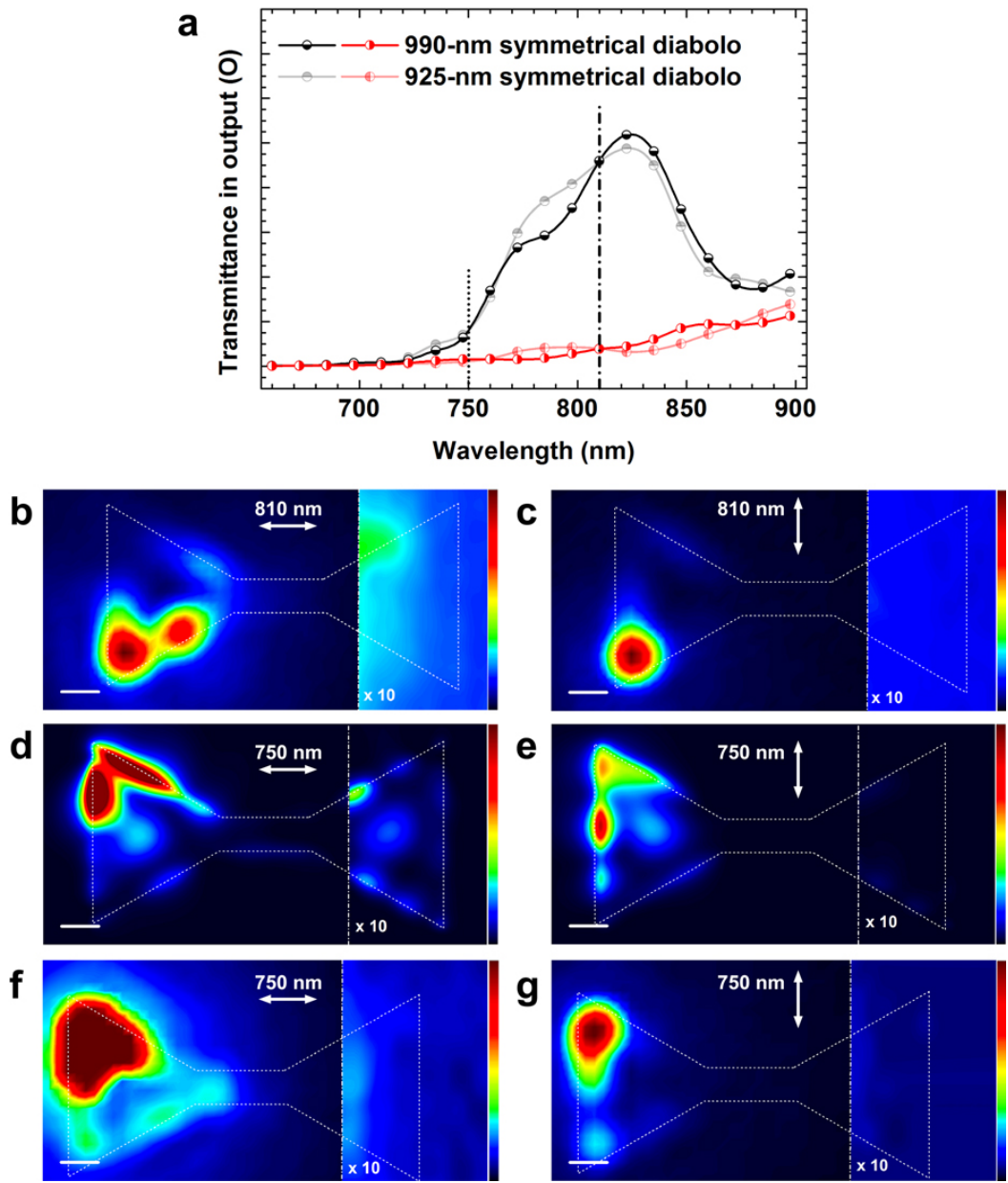

**Figure S10:** (a) Near-field transmittance spectra calculated at readout (O) for an excitation in (I) for the symmetrical diabolo with 990 nm sided triangular pads shown in Fig S1e. Horizontally (respectively vertically) split symbols correspond to the 0° (resp. 90°) polarization direction. The light-shaded spectra correspond to the symmetrical diabolo with 925-nm triangular pad studied in main text (See Fig. 5h). The dash-dotted and dotted lines indicate the experimental excitation wavelength (810 nm and 750 nm respectively). (b, c) Image plane nPL maps for the diabolo shown in Fig S1e obtained upon excitation in the lower left corner at 810 nm. (d, e) Transmittance maps for the diabolo shown in Fig S1e upon non-resonance excitation in (I) at 750 nm. (f, g) Image plane nPL maps for the diabolo shown in Fig S1e upon excitation in (I) at 750 nm. Scale bars are 200 nm. In maps in panels (b) to (g), the portion on the right of the dotted lines is plotted with a 10x magnified intensity using the same rainbow color scale.

The transmittance though mesoscopic symmetrical diabolo structures presents resonances in the visible and near-IR region as shown in Figure 6h. A small variation of the lateral size of the triangular pads has limited effect on the spectral characteristics of the transmittance modes. Indeed, in Figure S10a, the transmittance of the diabolo structures shown in Fig. 2a and Fig. S1e, which have the same channel size but a pad size difference of ca. 70 nm, is virtually identical. In particular, the transmittance curves coincide at 810 nm (vertical dash-dot line in Fig. S10a). Experimentally, the image plane nPL maps recorded on the larger diabolo show significant transmittance when the excitation is polarized along the longitudinal direction (Fig. S10b) but none when the polarization is orthogonal to the main axis (Fig. S10c) as already described for the smaller diabolo in the main text (See Fig. 2d, e).

The spectra suggest that the transmittance can be spectrally suppressed for both polarizations by exciting the structures out of the resonances peaks, for example at the wavelength of 750 nm (dotted vertical line in Fig. S10a) where the spectral intensity is incidentally the same for both structures and low for both polarization.

Accordingly, simulated near-field transmission maps for an excitation in the upper left corner at a wavelength of 750 nm yields no or very small transmittance in the output (O) region for either 0° (Fig. S10d) or 90° (Fig. S10e) excitation polarization.

Experimental image plane nPL upon excitation at 750 nm are displayed in panels (f) and (g) of Figure S10. The patterns in the excitation pad match closely the simulated near-field and none of the two polarization conditions yields measurable transmittance as observed in the output region plotted with a 10x magnified color scale.

This further confirms the delocalized mode-mediated transmittance in the resonant diabolo structures.